\documentclass{jpsj3}

\title{Behavior of the Quantum Critical Point and the Fermi-liquid Domain in the Heavy Fermion Superconductor CeCoIn$_{5}$ studied by resistivity
}

\author{Ludovic~Howald$^1$\thanks{E-mail address:ludovic.howald@gmail.com}, 
Gabriel~Seyfarth$^2$,
Georg~Knebel$^1$,
Gerard~Lapertot$^1$,
Dai~Aoki$^1$,
and Jean-Pascal~Brison$^1$
\thanks{E-mail address: jean-pascal.brison@cea.fr} %\\
}
\inst{$^1$CEA-INAC-SPSMS, 17 rue des Martyrs, 38054 Grenoble cedex 9, France\\
$^2$presently DPMC-Universit{\'e} de Gen{\`e}ve, 24 quai Ansermet, 1211 Gen{\`e}ve 4, Switzerland\\
}
\abst{We report detailed very low temperatures resistivity measurements on the heavy fermion compounds Ce$_{1-x}$La$_{x}$CoIn$_{5}$ ($x=0$ and $x=0.01$),  with current applied in two crystallographic directions [100] (basal plane) and [001] (perpendicular to the basal plane) under magnetic field applied in the [001] or [011] directions. We found a Fermi liquid ($\rho \propto T^{2}$) ground state, in all cases, for fields above the superconducting upper critical field. We discuss the possible location of a field induced quantum critical point with respect to $H_{c2}(0)$, and compare our measurements with the previous reports in order to give a clear picture of the experimental status on this long debated issue. 

}

\kword{CeCoIn$_5$,resistivity,$H_{c2}(0)$,Quantum Critical Point,Fermi Liquid}

\begin{document}
\maketitle

%***************************************************Introduction********************************************************************

\section{Introduction}

Quantum critical points (QCP) and their associated quantum phase transitions (QPT) are a major subject in strongly correlated electron systems, probably because they challenge our current understanding of competing ground states and are accompanied with the most dramatic effects of correlations like e.g. the strong mass renormalization in heavy fermion systems \cite{Lohneysen2007}. They are suspected to drive the occurrence of new states of matter like the breakdown of the topical Fermi liquid regime or unconventional superconducting states \cite{Mathur1998}. The ``115'' family \cite{Sarrao2007} is a central issue for this field, as it concentrates most of the theoretical questions connected with QCP together with a large panel of experimental probes (pressure, doping, field, neutron, thermodynamic and transport measurements, and others). However, after almost 10 years of studies, even some very basic experimental questions remain open, particularly for the case of CeCoIn$_{5}$. This system has a superconducting ground state at ambient pressure below 2.3 K, and shows no evidence of magnetic order in zero magnetic field. It is nevertheless commonly accepted that it is located close to a magnetic instability \cite{Sarrao2007} and strong antiferromagnetic spin fluctuations are observed in the normal state \cite{Kohori2001,Kawasaki2003,Sakai2010}. 
Furthermore, an antiferromagnetic phase has been observed under high magnetic fields in the basal plane, but only inside the superconducting mixed state \cite{Young2007, Kenzelmann2008, Koutroulakis2010,Ikeda2010}. %Antiferromagnetic ordering induced by the superconducting paramagnetic limitation was proposed for this region\cite{Ikeda2010}.
CeCoIn$_{5}$ is also easily turned antiferromagnetic upon Rh doping \cite{Goh2008}, as well as Cd\cite{Pham2006} and Hg\cite{Tokiwa2008} doping. And back in 2003, it was already discovered by two groups \cite{Paglione2003, Bianchi2003} that for fields parallel to the c-axis, the range of existence of the Fermi liquid regime seemed to shrink on approaching $H_{c2}(0)$ from the high field region, pointing to a possible QCP in the immediate vicinity of that field.

The aim of the present work is rather modest, but exemplifies the kind of difficulties encountered in this field : a precise determination of the phase diagram associated with the field induced quantum critical point in CeCoIn$_5$, as determined by charge transport. Indeed, since the work of Paglione et al \cite{Paglione2003}, it has been a matter of debate whether or not there is a QCP located exactly at $H_{c2}(0)$. The interest of that question is naturally directly related to the origin of the QCP (superconducting, magnetic, local, or others), the connection between the unusual behaviour of the upper critical field and the QCP, and the theoretical challenge to explain such a coincidence, see e.g. ref. \cite{Fujimoto2008}. It has been stressed \cite{Bianchi2002} that because the superconducting transition is first order, a QCP right at $H_{c2}(0)$ could not originate from the superconducting transition, but explaining the coincidence of a magnetic QCP with $H_{c2}(0)$ is by no way simple (see for example the discussion in ref. \citen{Ronning2005}). 

However, the present status of the experimental results is unclear : sticking first to the resistivity results, those of reference  \citen{Paglione2003} have been confirmed by Bianchi et al.\cite{Bianchi2003}, Ronning et al. (also for the field in the basal plane) \cite{Ronning2005},and further by Tanatar et al. \cite{Tanatar2007}. In ref. \citen{Tanatar2007}, Tanatar et al. confirmed their previous results also with resistivity measured along the c-axis. Doping the In site with Sn decreases the value of the upper critical field and appears to increase in a similar way the Fermi-Liquid domain \cite{Bauer2005}, which reinforces the previous statements. But the c-axis transport data of Malinowski et al. \cite{Malinowski2005} points to a critical field that would be well below $H_{c2}(0)$, contradicting these results, and in particular those of Tanatar et al. \cite{Tanatar2007}. A clear separation of the upper critical field and the extrapolated QCP has been observed at high pressure that would clearly locate an eventual QCP inside the superconducting domain.\cite{Ronning2006}

Switching to other probes, specific heat ($C_p$) is in principle an ideal thermodynamic reference to point out non Fermi liquid behaviour. From this respect, all measurements in the pure or Sn doped systems \cite{Bianchi2003, Bauer2005} indicate a ``crossover regime'' clearly above $H_{c2}(0)$, meaning that for fields slightly above $H_{c2}(0)$, no saturation of $C_p/T$ is observed, even when a $T^2$ regime is still measured on resistivity. Analysis of the specific heat from spin fluctuation theories nevertheless point to the proximity of the QCP to  $H_{c2}(0)$, with a continuously increasing logarithmic divergence on approaching the upper critical field \cite{Bianchi2003, Bauer2005} . However, a main difficulty with this probe is that at least 70\% of the bare signal at 100mK is coming from the hyperfine contribution, which severely limits the precision of the determination of the Fermi liquid border, particularly when it reaches very low temperatures (of order 0.1K or below). 

Thermal conductivity \cite{Tanatar2007} does not precisely locate the quantum critical point, but the observation of an anomalous ratio ``heat over charge transport conductivity'', that would remain above the Lorenz number down to the lowest temperatures, has been presented as an evidence for a breakdown of the quasiparticle concept for $H=H_{c2}(0)$, and a consequence of the presence of a QCP precisely at this field. 
Another thermal probe, the thermoelectric power \cite{Izawa2007}, is consistent with the proximity of a QCP at $H_{c2}(0)$, but cannot locate it precisely: the Seebeck coefficient $\nu/T\propto\gamma\propto A^{1/2}$ scales with specific heat and the $A$ coefficient of resistivity, and is strongly increased in the proximity of the superconducting transition. Another probe which has been used is Hall effect \cite{Singh2007}: in fact, this probe cannot by itself give a criterium of Fermi liquid behaviour, but it allows to single out a caracteristic field, function of temperature, which happens to coincide with the line drawn from resistivity data. The interest of the Hall determination is that it would be less prone to magnetoresistance effects which affect resistivity at the lowest temperatures (or the highest field), allowing for a more precise determination of the border of the Fermi liquid domain close to $H_{c2}(0)$. Hall effect would clearly locate the putative QCP below $H_{c2}(0)$\cite{Singh2007}.

The superconducting transition is first order at low temperature high field\cite{Bianchi2002}, which exclude the possibility of a superconducting quantum critical point. Furthermore, the superconducting coherence length in CeCoIn$_5$ is large ($>50$\AA). This implies that superconducting fluctuations in the paramagnetic phase are expensive in terms of energy and therefore unlikely to be present in wide regions of the phase diagram as observed.

So there seems to be a larger number of experimental facts pointing to a QCP right at $H_{c2}(0)$, even though contradictions remain and precision of all measurements seems not sufficient to give a clearcut answer on the boundary of the Fermi liquid regime according to the criterium of the probe. The present work therefore endeavoured to improve on the resistivity measurements combining:
\begin{itemize}
\item low temperature transformer (gain 1000) and extended temperature measurement (down to $8$ mK), in order to reduce the noise and work at low enough currents to prevent heating down to the lowest temperature measured.
\item c-axis transport, which has been pointed out as the most sensitive to Fermi liquid regime breakdown \cite{Tanatar2007}, and has the advantage of strongly limiting magnetoresistance effects (configuration of field parallel to the current), together with the more usual \cite{Paglione2003,Bianchi2003,Bauer2005} a-axis transport for detailed comparison to previous work.
\item a slightly La doped sample (about $1\%$ doping), (current also along the c-axis) to further damp magnetoresistance effects, strongly limited by impurity scattering.
\item two different fields orientations (parallel to c-axis or at $45\,^{\circ}$ from c-axis), to check the coincidence of the QCP and $H_{c2}(0)$ on the same samples, during the same experiment.
\end{itemize}

With all these improvements, we can exclude that resistivity points to a QCP pinned at $H_{c2}(0)$. Instead, like in the Hall measurements \cite{Singh2007} or the older c-axis transport measurements  \cite{Malinowski2005}, the critical field $H_{QCP}$ (if any) would be well inside the superconducting mixed phase. Due to superconductivity, the criticality at $H_{QCP}$ is not directly experimentally accessible.

%***************************************************Experience********************************************************************

\section{Experimental details}

We have measured the resistivity with a standard 4 wires technique down to 8 mK and in magnetic fields up to 8.5 Tesla on three different samples. In all measurements,  we used a low temperature transformer with gain 1000, and a low noise preamplifier (built in CNRS-N\'eel Institut) to amplify the signal. The current at lowest temperature was carefully adjusted to avoid heating of the samples. Samples A and B have pure CeCoIn$_{5}$ composition, sample C is 1\% doped with Lanthanum on the Cerium site (Ce$_{0.99}$La$_{0.01}$CoIn$_{5}$). The samples were cut in a bar shape of sizes about $0.2\times0.2\times0.5mm^3$ for samples B and C and about $1.2\times0.2\times0.2mm^3$ for sample A.  Resistivity was measured with current applied in the basal plane (a-axis) for sample A and along the c-axis, for the samples B and C. 
Contact resistances smaller than $1 m\Omega$ are necessary for the voltage leads, to take full advantage of the low temperature transformer. To achieve such small contact resistances, gold stripes (with Ti underlayer) were deposited under vacuum, after ion gun etching of the surface. On these stripes $38\mu m$ gold wires are fixed by spot welding. The three samples were glued with a small amount of G.E. varnish (enough to prevent grounds)  directly on a silver plateau which is screwed on an Attocube$\textsuperscript{\textregistered}$ piezzo-rotator. This allows in-situ change of field orientation. The angle is determined by a Hall probe. 

\begin{figure}[hbt]
\centering
\includegraphics[width=0.9\textwidth]{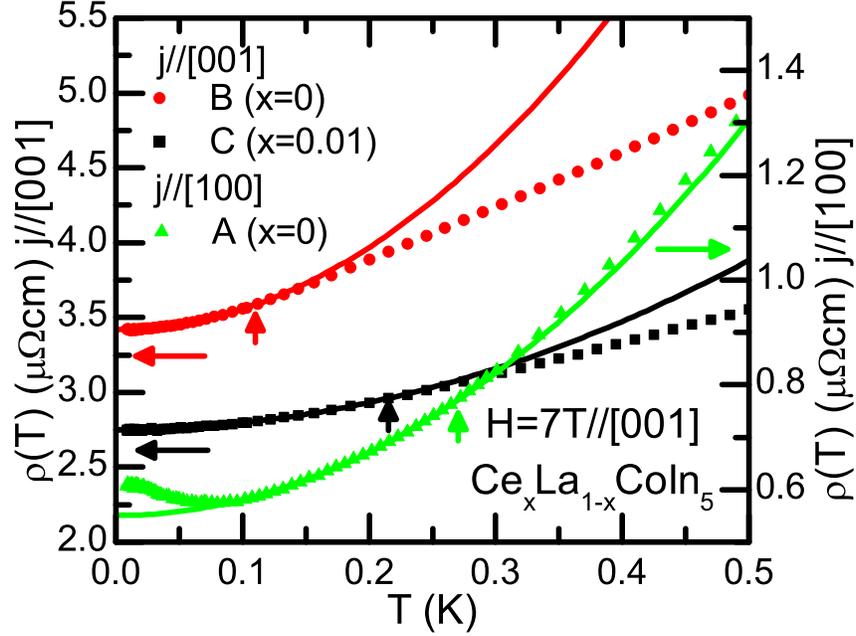}% Here is how to import EPS art
\caption{(color online) Typical resistivity curves for the three samples. When magnetic field and current are applied perpendicularly, magneto-resistance effects are strong at low temperatures ($\omega_{c}\tau>>1$, sample A). In sample B these effects are strongly reduced as the magnetic field and current are parallel. The same is true for sample C, in which $\omega_{c}\tau$ has been further decreased through a light (1\%) La doping. Arrows indicate the upper limit of the Fermi liquid regime. }
\label{resData} 
\end{figure}

Figure \ref{resData} shows a typical measure $\rho(T)$ for $H = 7$ T  applied along the [001] direction for the three samples. In sample A we see a clear upturn at low temperatures. This upturn increases with magnetic field, which is an indication that the behavior of the quasi-particles is dominated by magneto-resistance effects when entering the quantum regime $\omega_{c}\tau>>1$. Therefore, as in previous works, we have to discard the lowest temperature data, and we can fit the resistivity with a Fermi-liquid law ($\rho=\rho_0+AT^2$) only above about 80mK in this sample. This is not the case for samples B and C, for which magnetic field and current are parallel (H$\parallel$c$\parallel vec{j}$). In sample C, La impurities decrease further the mean free path $\tau$ and therefore the magneto-resistance effects. The residual resistivity shows a weak field dependence with two different slopes: positive on sample B and negative on sample C. At H=0 we can extrapolate the residual ratio of the three samples $RRR = \rho(300K)/\rho(T\rightarrow0)$: RRR(Sample A) $\simeq$ 335, RRR(Sample B) $\simeq$ 24 and RRR(Sample C) $\simeq$ 20. Variation between samples B and C reflects the presence of La impurities. The high value of the RRR of sample A demonstrates the high quality of our samples.

\begin{figure}[hbt]
\centering
\includegraphics[width=0.9\textwidth]{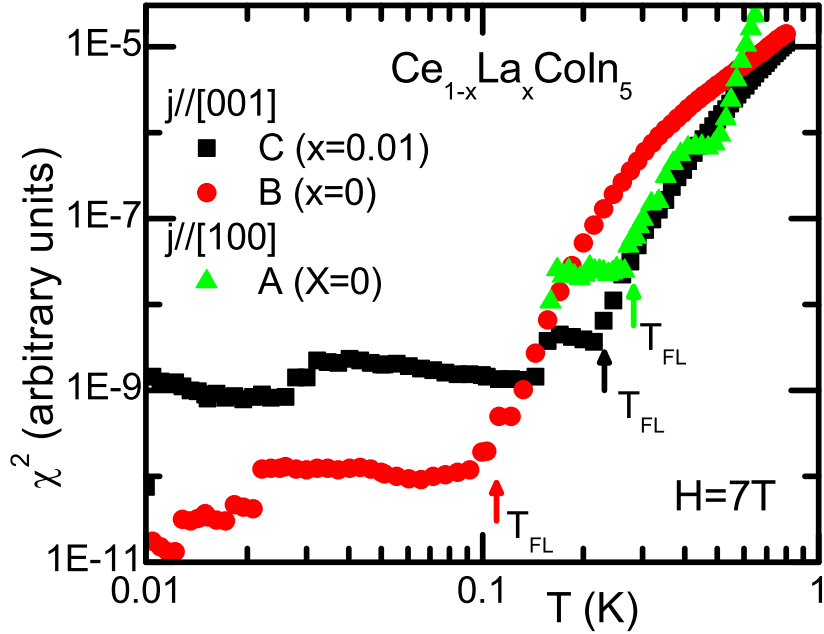}% Here is how to import EPS art
\caption{(color online) To determine the domain in which the data can be fitted with a Fermi-liquid law $\rho (T)=AT^2+\rho_{0}$, we plot the deviation from this law measured by the ``chi-square'' value $\chi^2$ and normalized by the number of points. For $T<T_{FL}$ the value is constant and corresponds to the noise of the measurement; at higher temperatures, the ``chi-square'' value increases exponentially  due to systematic deviations. Data are shown for the three samples at $H=7$ T $H\parallel$ [001]. The small steps, at very low temperatures, correspond to changes of gain or sensibility in the lock-in amplifier used for the measurement, not perfectly calibrated.}
\label{Xi2}
\end{figure}

Figure \ref{Xi2}, shows how we determine the upper boundary of the Fermi liquid regime $T_{FL}$. The curves are the ``chi-square'' values $\chi^2(T)$ between the data points and a fit of these points ($\rho (T)=AT^2+\rho_{0}$), from the lowest temperature up to temperature $T_{FL}$. The $\chi^2(T)$ function is normalized by the number of points in the fitted interval. For sample A the lowest value taken into account is 80mK due to low temperature magneto-resistance effects mentioned previously, whereas we could go down to $8mK$ for the two other samples. At low temperatures, $\chi^2$ is roughly constant, and its value reflects white noise on the data points. Above $T_{FL}$, $\chi^2$ increases logarithmically due to systematic deviations. 

\begin{figure*}%two column image
\centering
\includegraphics[width=\textwidth]{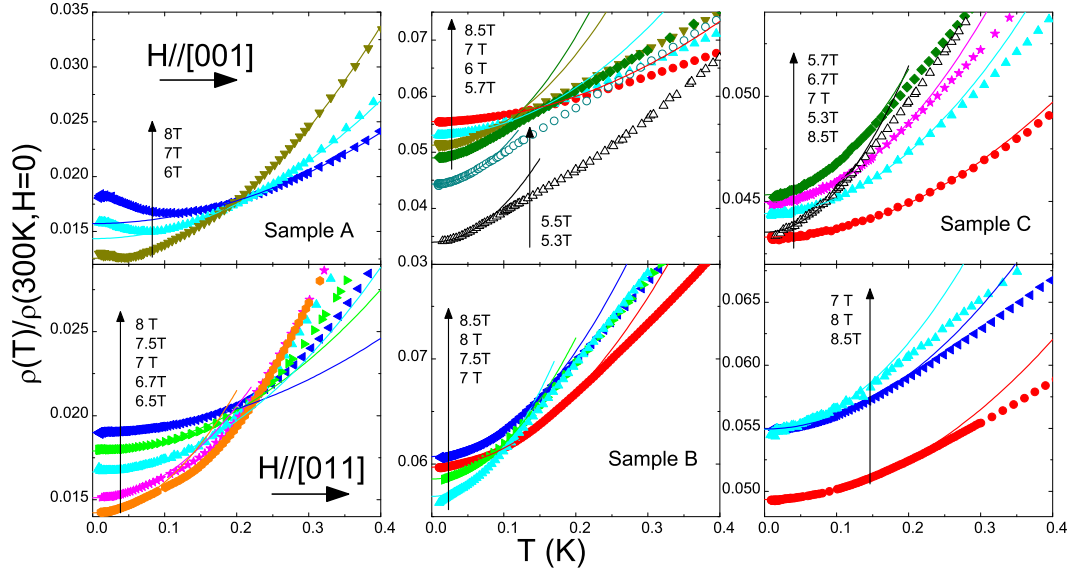}% Here is how to import EPS art
\caption{Various measurements $\rho(T)$ at a constant magnetic field (raw data). Each column is dedicated to one sample, each row, to a given field orientation (upper row, $H\parallel [001]$, lower $H\parallel [011]$). The same colours and symbols are used for each field value on all graphs.}
\label{AllRes}
\end{figure*}

%***************************************************Results********************************************************************
\section{Results}

\begin{figure}
\centering
\includegraphics[width=0.7\textwidth]{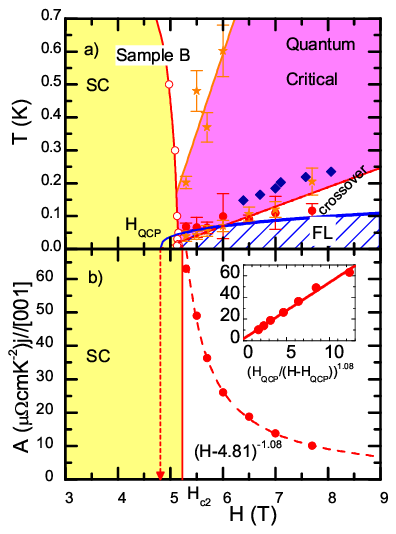}% Here is how to import EPS art
\caption{ (color online). (a): phase diagrams for sample B CeCoIn$_5$, $H\parallel [001]\parallel \vec{j}$. Yellow: superconducting phase, dashed blue: Fermi-liquid domain from ``chi-square'' analysis. Red vertical dotted lines, position of the QCP, deduced from the divergence of the $A$ coefficient of resistivity. The quantum critical region (pink), where resistivity is linear in temperature,  (same ``chi-squar'' analysis) points to the same field. To vanish at the same value, the $T_{FL}$ line must follow $T_{FL} \propto (H-H_{QCP})^{z/2}$ with $z<2$. Blue diamonds are the values obtained from anomaly in Hall effect measurements \cite{Singh2007}.
(b) Divergence of the $A$ coefficient, fitted with a law $A\propto \frac{1}{(H-H_{QCP})^{-\alpha}}$, $\alpha=1.08$: it points to a value $H_{QCP}<H_{c2}$. Inset shows the validity of the law.
}
\label{Gra_001b}
\end{figure}

\begin{figure}
\centering
\includegraphics[width=0.7\textwidth]{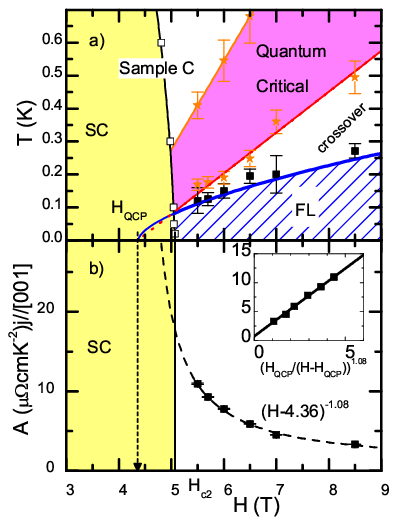}% Here is how to import EPS art
\caption{ (color online). (a) phase diagrams for sample C Ce$_{0.99}$La$_{0.01}$CoIn$_5$, $H\parallel [001]\parallel \vec{j}$. Yellow: superconducting phase, dashed blue: Fermi-liquid domain from ``chi-square'' analysis. Black vertical dotted lines, positions of the QCP, deduced from the divergence of the $A$ coefficient of resistivity. The quantum critical region (pink), where resistivity is linear in temperature,  (same ``chi-square'' analysis) points to the same field. To vanish at the same value, the $T_{FL}$ line must follow $T_{FL} \propto (H-H_{QCP})^{z/2}$ with $z<2$. 
(b) Divergence of the $A$ coefficient, fitted with a law $A\propto \frac{1}{(H-H_{QCP})^{-\alpha}}$, $\alpha=1.08$: it points to a value $H_{QCP}<H_{c2}$. Inset shows the validity of the law.
}
\label{Gra_001c}
\end{figure}

Figure \ref{AllRes} shows fits for the various samples and magnetic fields. Previous similar studies \cite{Bianchi2003,Paglione2003} were done with the geometry of sample A (magnetic field [001] applied perpendicular to the current [100]). Our measurements of sample A agree with previous reports, but the deduced value of $T_{FL}$ has a large uncertainty owing to the restricted temperatures range imposed by the low temperature magneto-resistance effects (between $80$ mK and $T_{FL}$). Moreover, $T_{FL}$ is strongly dependent on the lowest temperature used for the fit. We believe that the resulting $T_{FL}$ obtained with this configuration is slightly overestimated. 

Figures \ref{Gra_001b}b and \ref{Gra_001c}b show the field dependence of the $A$ coefficient of resistivity ($\rho(T) = \rho_0 + AT^2$) for samples B and C. The strong increase  of the $A$ coefficient reported on figures \ref{Gra_001b}b and \ref{Gra_001c}b is commonly found on approaching a QCP, and we defined a critical field H$_{QCP}$ by fitting A as proportional to $(H-H_{QCP})^{-\alpha}$.

The power of the divergence $\alpha$ can be obtained by fitting the 5 sets of $A$ values simultaneously: we measured  3 samples in 2 fields orientations, but we discarded the data of sample A - H$\parallel$[001], for the above mentioned reasons. %Figures \ref{Gra_001_All}c and \ref{Gra_45deg}b not shown for sample C. 
More precisely, we have allowed for a regular contribution ($A0$) to the A coefficient: $A_i^\theta=(A0)_i^\theta+(A1)_i^\theta((H-H_{QCP})/H_{QCP})^{-\alpha}$, $i=[A,B,C]$, $\theta=[0\,^{\circ},45\,^{\circ}]$ with the same exponent $\alpha$ for all fits, and $H_{QCP}$ depending only of sample composition and field orientation (same $H_{QCP}$ for samples A and B). $(A0)_i^\theta$ have the constraint to be positive as they are the values of $A$ far from criticality, and they are found to be very small (could even be forced at $0$) except for sample C. The exponent is found very close to one ($\alpha=1.08\pm 0.6$).

\begin{figure}[htbp]
\centering
\includegraphics[width=1\textwidth]{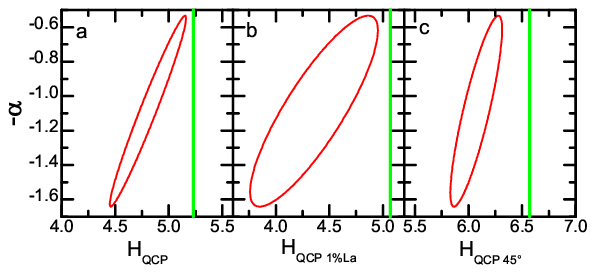}% Here is how to import EPS art
\caption{(color online) Confidence region boundary (one standard deviation) from the fit of the divergence on the $A$ coefficient of resistivity, of the two parameters $\alpha$ and $H_{QCP}$: a) for sample B with H$\parallel$c-axis, b) for sample C with H$\parallel$c-axis and c) for samples A and B with field applied at 45$\-,^{\circ}$ to the c-axis. In the three cases, the divergence of $A$ locates $H_{QCP}$ below $H_{c2}(0)$ (green vertical line).}
\label{GraElipse} 
\end{figure}

We can use the divergence of $A$ to extrapolate the location of the QCP on the magnetic field axis. Such a divergence is expected for example in the spin fluctuation scenario \cite{Moriya1995}. 
Figure \ref{GraElipse} shows the confidence region boundary (one standard deviation) between the exponent $\alpha$ and each $H_{QCP}$ for the three different geometries. In the three cases, the divergence of $A$ locates the QCP at fields below $H_{c2}(0)$. Also, at $H_{c2}(0)$ we can still fit resistivity with a $T^2$ law up to about 50mK in the pure sample (B), and up to about 100mK for the doped one (C): figures \ref{Gra_001b}a and \ref{Gra_001c}a.
These two observations clearly indicate that resistivity does not point to a field induced QCP precisely at $H_{c2}(0)$, as previously stated for this compound \cite{Bianchi2003,Paglione2003}. Instead, they confirm the phase diagram for Fermi-liquid domain suggested from Hall effect measurements\cite{Singh2007}. This is also confirmed with the data analysis of sample C (figure \ref{Gra_001c}a), which has more disorder and is therefore less prone to magnetoresistance effects at low temperature/high fields: if a QCP should exist at ambient pressure in this compound, it is hidden by the superconducting phase.

\begin{figure}[htbp]
\centering
\includegraphics[width=0.7\textwidth]{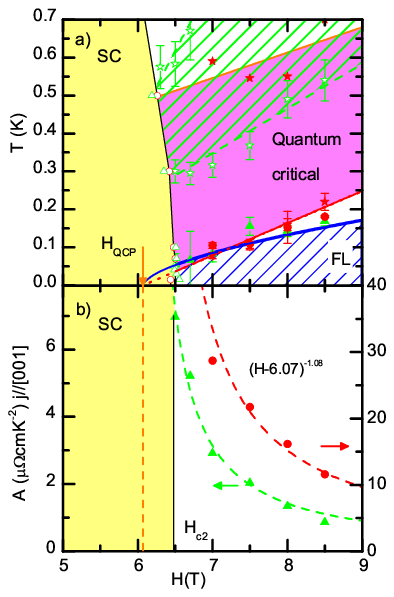}% Here is how to import EPS art
\caption{ (color online) When the magnetic field is applied along $H\parallel$[011] direction, the magneto-resistance effects are weak for all the samples.  Sample A (j//a-axis) full green triangles: Fermi liquid and A term, open green triangles: $H_{c2}$ and open green stars: $T$-linear (quantum critical) regime. Idem for sample B (j//c-axis) with red circles and full red stars. (a) The Fermi-Liquid domain is the same for samples A and B, contrary to the $T$-linear regions (stars) (see text). (b) Divergence of the $A$ coefficient ($A\propto \frac{1}{(H-H_{QCP})^{-\alpha}}$, $\alpha = 1.08$) for both samples, pointing to the same $H_{QCP}$.
}
\label{Gra_45deg}
\end{figure}

We can also compare the value of field for which we extrapolate a divergence of the $A$ coefficient, with the value of the field for which we extrapolate that $T_{FL}$, the upper bound of the Fermi-liquid regime, goes to zero. 
In the Hertz-Millis scenario \cite{Millis1993}, the Fermi-liquid border is linear in the parameter controlling the approach of an antiferromagnetic QCP. If we identify this parameter with the magnetic field, we expect: $T_{FL} \propto (H-H_{QCP})^{z/2}$, with $z=2$ the dynamical exponent. A simple examination of figures \ref{Gra_001b}a, \ref{Gra_001c}a and \ref{Gra_45deg}a shows that linear extrapolation of $T_{FL}$ always yields a value for $H_{QCP}$ much lower than that extrapolated from the divergence of $A$. It further confirms that if there should be a QCP in CeCoIn$_5$, it has to be located below $H_{c2}(0)$. However, the discrepancy of $H_{QCP}$ as deduced from the divergence of $A$ or the linear extrapolation $T_{FL}(H)=0$ is problematic. 

Another point of view would be to fix the value of $H_{QCP}$ from the divergence of $A$, and then consider $z$ as an adjustable free parameter: for example, for $H_{QCP} = 4.81$ Tesla (sample B), obtained with $\alpha = 1.08$, we then obtain $z=0.7 \pm 0.14$. Similarly, we get  $z=1.24\pm 0.12$ for the doped sample C, however, in the latter case, extrapolation of $H_{QCP}$ from the field divergence of $A$ is quite arbitrary as the increase of $A$ down to $H_{c2}(0)$ remains very modest. When magnetic field is applied at 45$\,^{\circ}$ from c-axis, we get $z=1.22\pm 0.08$ for samples A and B. Note that even if we leave the value of $H_{QCP}$ vary in the full confidence interval obtained from the divergence of the $A$ coefficient, we cannot obtain $z=2$ for the dynamical exponent. We found $z \in [0.4;0.9],[0.9,1.6],[0.5;1.4]$ respectively for the three cases discussed previously. Conversely, a value of $z=1$ is expected in a scenario where the f-electron do not form bands \cite{Reyes2009}. However, a main difficulty might be that the standard scenario, which does not consider polarisation of the bands under field, is simply not applicable to a field induced QCP. Experimentally, the mechanism driving the destruction of the AFM order under field, without any kind of metamagnetic transition is also unclear. For CeCoIn$_5$, one might argue that the jump of magnetization observed a $H_{c2}(0)$ \cite{Tayama2002} is not only a diamagnetic jump, but has also a paramagnetic origin \cite{Kos2003}, which could reflect such a metamagnetic transition. In any case, quantitative theoretical prediction is missing for such field induced QCP. Let us note however, that the factor two we find between the power law for the divergence of the $A$ coefficient and the field dependence of $T_{FL}$, is what is simply expected from a dimensional point of view, if the approach of the QCP is governed by the collapse of a single energy scale ``$T_0$'' : $\rho\approx A'(T/T_0)^2 \Longrightarrow A\approx (1/T_0)^2$, and $T_{FL}\propto T_0$, independently of the identification of $T_0$ with a Kondo temperature, spin fluctuation temperature...

Another way to delimitate the Fermi-liquid regime is to observe the growth of the non-Fermi liquid behaviour: a $T$-linear regime is often observed above the $T^2$ regime, and has been reported in the first studies of CeCoIn$_5$ \cite{Petrovic2001,Paglione2003,Ronning2006,Tanatar2007}. With the same technique than for the Fermi-liquid region, we determine the $T$-linear domain (pink region in figures \ref{Gra_001b}a and \ref{Gra_001c}a). The lower bound of the $T$-linear domain matches that of the observed dip in the differential Hall coefficient associated with departure from Fermi-liquid regime\cite{Singh2007}. The onset temperature of this $T$-linear regime of the resistivity is extrapolated to vanish at the same magnetic field value ($H_{QCP}$) than the divergence of the $A$ coefficient (see figures \ref{Gra_001b}a, \ref{Gra_001c}a and \ref{Gra_45deg}a). This is consistent with a linear behaviour of the resistivity down to $T=0$, if it could be measured at $H_{QCP}$.

Figure \ref{Gra_45deg} shows the same analysis when the magnetic field is applied in the direction [011]. In this case, magneto-resistance effects are weak enough whatever the direction of the applied current, and therefore we can compare the two pure samples A ($\vec{j}\parallel$[100]) and B ($\vec{j}\parallel$[001]). As in the previous case, both the divergence of the $A$ coefficient of resistivity and the collapse of the Fermi-liquid domain happen inside the superconducting phase. It is interesting to point out, that even if the absolute values of resistivity are very different, the two Fermi-liquid borders $T_{FL}$ and divergences of $A$ coefficient coincide for both samples. This is a good indication of the validity of our analysis. However, the $T$-linear domains have different borders depending on the current direction. This underlines an intrinsic difficulty of discussing Fermi liquid borders from transport measurements. Indeed, the Fermi liquid domain is inherently an isotropic property of the system, which should not depend on the current direction. But if the breakdown of the Fermi liquid regime is associated with singularities located on some peculiar regions of the reciprocal space (it has been suggested that for CeCoIn$_5$, quasiparticles disappear along the c-axis \cite{Tanatar2007}), it will affect differently transport depending on the current direction, which can easily lead to different determinations of ``non Fermi liquid'' behaviour. In the present case, these measurements for $H\parallel[011]$ confirm that the c-axis is much closer to ``criticality'' than the a-axis, when using a criterium of $T$-linear behaviour.

%***************************************************Discussion********************************************************************
\section{Discussion}

Our analysis of temperatures dependences of the resistivity converge to a QCP located clearly below $H_{c2}(0)$. This is in good agreement with previous measurements of specific heat, Hall effect and thermal expansion of other authors \cite{Singh2007,Bianchi2003,Donath2008}, that point to a QCP located inside the superconducting dome. 

Nevertheless, as for other heavy fermion systems, it is difficult to go beyond this qualitative analysis, and deduce more quantitative information on the nature of the QCP from the precise laws and exponents of the divergence of $A$ coefficient or field variations of $T_{FL}$. Spin fluctuation models do not predict a divergence of the specific heat for antiferromagnetic fluctuations (at $T\rightarrow 0$), whereas it does predict a divergence of the $A$ coefficient of resistivity. Experimentally, a diverging behaviour of both quantities is observed in the measured temperature range. However, specific heat measurements cannot be performed below 80mK, so they remain compatible with any scenario. Scaling even matches prediction of spin fluctuation models, as saturation of the specific heat is only expected at very low temperature close to a QCP. 
A problem with the spin fluctuation model is that it does not predict the $T$-linear regime observed in resistivity. This has triggered the theoretical development of so called ``unconventional models'' of criticality, like a breakdown of the Kondo effect which could generate a divergence of the specific heat, and predict the $T$-linear dependence of resistivity, at the expense of a change of the Fermi surface. Presently, such a Fermi surface change has not been observed in CeCoIn$_5$, despite the rare possibility to perform de Haas van Alphen experiments below $H_{c2}$. In any case, there are still very few quantitative predictions of these new models that we could test with the present experiment.

Curiously, our data match several predictions of the phase diagram proposed by A. Rosch \cite{Rosch2000}, for an anti-ferromagnetic induced QCP with magnetic impurities (and for small effect of the magnetic field). For example, the authors of Ref. \citen{Rosch2000} predict two different behaviours :  $T_{FL} \propto (H-H_{QCP})^{1/2}$ and $T_{Linear} \propto (H-H_{QCP})$, very close to our experimental observations. This may seem at odds with the well known high quality of single crystals available for this system, however, from the ``magnetism point of view'', there is a clear ``smoking gun'' for the presence of unusual magnetic disorder in CeCoIn$_5$ : for example, the unusually large specific heat jump at the superconducting transition \cite{Rosch2000,Petrovic2001}, and the jump of magnetization observed at $H_{c2}$ even close $T_c$ \cite{Ikeda2001} can be explained by the presence of magnetic disorder like remaining fluctuating paramagnetic centers \cite{Kos2003}. A complete and quite successful model for the appearance of coherence in this system (in the framework of ``Kondo-Lattice physics'') has also been proposed, which points to residual ``uncondensed'' Kondo impurity centers in CeCoIn$_5$ down to very low temperatures \cite{Nakatsuji04}.

A possible way to have our data of $T_{FL}(H)$ not contradict the linear behaviour of the Hertz-Millis scenario, could be to claim that no QCP is present in the ($H,T,P=0$) phase diagram. If the QCP is located at negative pressure, then in the plane P=0 of phase space, the Fermi liquid boundary ($T_{FL}(H)$) would be an hyperbola. This cannot be excluded by our measurement of $T_{FL}$, because superconductivity hides the low field regime. However, the apparent divergence of the A coefficient at a finite field seems unlikely in such a scenario.

% ***************************************************  compare two compounds **********************************************************

Another approach would be to compare CeCoIn$_5$ with other prototypes of quantum critical points, and particularly of field induced quantum critical points. From this point of view, probably the best documented case is that of YbRh$_{2}$Si$_2$: a divergence of the $A$ coefficient, an anomalous $T$-linear behaviour of the resistivity at $H_{QCP}$, together with a well identified antiferromagnetic ordered phase have been reported\cite{Custers2003}. The Gr\"uneisen ratio in the critical region has the same temperature dependence for the two compounds \cite{Donath2008} and has been claimed to proof a ``local'' scenario\cite{Si2001} for the non Fermi liquid behaviour and QCP in YbRh$_{2}$Si$_2$.  This is also suggested by Custers et al. \cite{Custers2003}, in order to explain the exceptional broad range of existence of a linear temperature behaviour of the resistivity. It has also been stressed that recent experiments \cite{Friedemann2009} using Ir or Co doping of this system, support such a local scenario because they show that transport anomalies are not pinned to the magnetic QCP. In any case, even in the pure system, Knebel et al \cite{Knebel2006} had already shown that no true divergence of the $A$ coefficient was observed at the magnetic QCP, and that the range of observation of the $T^2$ law remains finite in the whole temperature-field phase diagram. 
It is interesting to note that comparison can be pushed a step further when looking at the ``critical exponents'' of YbRh$_{2}$Si$_2$ (data of Ref. \citen{Knebel2006}): in both cases, the divergence of the $A$ coefficient can be well fitted by a simple law : $A\propto (H-H_{QCP})^{-\alpha}$, and the dynamical exponent for $T_{FL} \propto (H-H_{QCP})^{z/2}$. The exponents are found to vary in the interval $\alpha \in [0.4;1.25]$ and $z \in [1.1;1.6]$ surprisingly similar to the case of the doped Ce$_{0.99}$La$_{0.01}$CoIn$_5$ sample, and also in contradiction with the Hertz-Millis scenario.

Whether these similarities originate in a similar mechanism for the QCP remains to be investigated. But a major interest of the case of CeCoIn$_5$ is that it combines the rare advantages of high purity crystals and a field scale ($H_{QCP}\approx 5T$) large enough for Fermi surface studies. Of course, de Haas-van Alphen studies in the superconducting phase are notoriously difficult, but they are possible in this system, meaning that both sides of the putative QCP can be probed \cite{Settai2001}. Up to now, they did not reveal any change as expected in the local scenarios of Kondo breakdown, but CeCoIn$_5$ might be a good candidate to test the most dramatic predictions of this class of QCP models, and so, worth a deeper look.

\section{Conclusion}

Fit of resistivity data down to very low temperatures (8mK) following $\rho(T)=\rho_{0}+AT^{2}$ allows us to determine the boundary of the Fermi liquid domain in CeCoIn$_5$ in the neighbourhood of $H_{c2}(0)$, for $H\parallel c$, with unprecedented precision. $T_{FL}$ does not vanish at $H_{c2}(0)$ in CeCoIn$_5$, and if a quantum critical point exists in this system, its location is at lower magnetic field and therefore hidden by the superconducting state: more precisely, a real quantum critical point may exist in this system within the superconducting phase, but he would then result from a deep interaction between superconducting and magnetic order (as proposed for example in ref.\cite{Ikeda2010}), with no direct relation with the non Fermi liquid behaviour observed above $H_{c2}$. Indeed, the non-Fermi liquid regime can only be sensitive to the magnetic fluctuations, owing to the very small size of the superconducting critical regime, and the first order nature of the superconducting transition at low temperatures in this compound. Moreover, no peculiar anomaly has been found for transport along the c-axis, meaning that a Fermi liquid regime is still observed on resistivity down to $H_{c2}(0)$ along this direction, albeit in a very restricted temperature range (below $50mK$). This is also confirmed by an accurate determination of the ``divergence'' of the A coefficient of resistivity. Moreover, the field dependence of $A$ and $T_{FL}$ are compatible with a QCP governed by the collapse of a single energy scale. We can explain the difference with some of the previous work, as due to improved precision and/or the use of a more favourable setup geometry, less prone to low temperature magneto-resistance effects. This may help to clarify the relationship between QCP and superconductivity in this compound, however, it also stresses the need for theoretical studies and predictions for field induced QCP.

\section*{Acknowledgment}

We would like to thank J. Flouquet for stimulating discussions. This work was supported by the French ANR grants SINUS and DELICE.

\bibliographystyle{jpsj}

\end{document}